\documentclass[prl,twocolumn,superscriptaddress]{revtex4}

\usepackage{amsfonts}
\usepackage{amsmath}
\usepackage{bbm}
\usepackage{graphicx}
\usepackage{bbold}
\usepackage{amssymb}
\usepackage{color}
\usepackage{subfigure}

\usepackage{mathtools}

\begin{document}

\title{Comment on ``Lack of a genuine time crystal in a chiral soliton model" \\by Syrwid, Kosior, and Sacha}

\author{Patrik \"Ohberg}
\affiliation{SUPA, Institute of Photonics and Quantum Sciences, Heriot-Watt University, Edinburgh EH14 4AS, United Kingdom}
\author{Ewan M. Wright}
\affiliation{SUPA, Institute of Photonics and Quantum Sciences, Heriot-Watt University, Edinburgh EH14 4AS, United Kingdom}
\affiliation{Wyant College of Optical Sciences, University of Arizona, Tucson, Arizona 85721, USA}

\maketitle

In a 2019 paper we proposed a model for a quantum time crystal based on a chiral soliton on a ring \cite{OW19}. Syrwid, Kosior, and Sacha (SKS) subsequently published a comment claiming that our system cannot realize a genuine time crystal \cite{SKS_comment}, and we responded to their comment \cite{OW_response}.  SKS have now published a paper with an extended version of their arguments \cite{SKS_lack} and the present comment is our response. We stand by our claim that our system can realize a quantum time crystal.

We state that we do not dispute the theoretical analysis presented by SKS, but rather our differences lie in some conditions built into their analysis which drive their final conclusions.  In particular, SKS consider the center-of-mass (COM) momentum or velocity of the chiral soliton to be a continuous classical variable not subject to any quantization conditions imposed by the ring \cite{SKS_lack}.  This would seem to be inconsistent with basic quantum mechanics: If one considers the Schr\"odinger equation for a system of $N$ bosons on a ring with two-body interactions, then the (COM)  motion is equivalent to that for a free-particle on the ring, complete with quantization of the COM momentum \cite{Bloch}.  But this assumes zero-temperature, and infinite associated thermal de Broglie wavelength, and SKS consider the thermodynamic limit where the COM momentum does indeed approach a continuous variable \cite{SKS_lack}.  So within the conditions set by SKS one can conceive of a chiral soliton that is well localized on the ring, and with a thermal de Brogile wavelength that is longer than the spatial extent of the chiral soliton but considerably shorter than the ring circumference. In this macroscopic limit the analysis of SKS is applicable and a genuine time crystal is not possible, the lowest energy occurring for zero COM momentum \cite{SKS_comment,SKS_lack}. We do not dispute this.

The problem is that we have never advocated using the above conditions.  In the conclusions of our initial publication we explicitly mention not relying on the thermodynamic limit \cite{OW19}.  More specifically, one can consider {\it mesoscopic rings} involving small atomic Bose-Einstein condensates whose thermal de Broglie wavelength can be substantially greater than the ring circumference.  In this case quantization of the COM velocity must be imposed as advocated in our previous papers \cite{OW19,OW_response}, and this is what can lead to the ground state arising for a non-zero COM momentum.  In addition, contrary to the statement by SKS \cite{SKS_lack}, we never claimed that the time crystal was formed using a strongly localized chiral soliton. A weakly localized chiral soliton that extends around the ring is also perfectly acceptable.  Once the restrictive conditions imposed by SKS are relaxed a quantum time crystal can be realized in our system as illustrated in our previous response \cite{OW_response}.

Finally, in the paragraph below Eq. 2 of their paper \cite{SKS_lack} SKS point out that the limit $N\rightarrow\infty$ is required to eliminate quantum fluctuations in the COM position of the rotating mean-field soliton, so that the system can persist indefinitely and give rise to what they term a genuine quantum time crystal.  But surely this does not encompass the experimental study of quantum time crystals of the type considered here, which by necessity involve a Bose-Einstein condensate with finite $N$ along with a finite lifetime due to intrinsic and experimental limitations.  Moreover, the physically essential process of monitoring the many-body quantum system will render it an open system, giving another source of a finite lifetime.  The point is that as long as the time for quantum fluctuations of the COM position to develop is longer than the system lifetimes then the requirement $N\rightarrow\infty$ is moot. In proposing our chiral soliton model for a quantum time crystal we tacitly assumed that it could be realized in the laboratory for a mesoscopic ring without recourse to the thermodynamic limit, albeit with a finite lifetime.

In conclusion, we resolutely stand by our claim that our system can realize a quantum time crystal.

\end{document}